\documentclass[useAMS,usenatbib]{mn2e}
\usepackage{graphicx}
\def\apj{ApJ}
\def\mnras{MNRAS}

\def\swift{{\it Swift~}}

\title[Constraint on dark energy from GRBs]
{Constraints on $w_0$ and $w_a$ of Dark Energy from 
High Redshift Gamma Ray Bursts}
\author[R. Tsutsui et al.]
{Ryo Tsutsui$^{1}$\thanks{E-mail: tsutsui@tap.scphys.kyoto-u.ac.jp (RT)},
Takashi Nakamura$^{1}$, Daisuke Yonetoku$^{2}$,Toshio Murakami$^{2}$,
\newauthor Sachiko Tanabe$^{2}$, Yoshiki Kodama$^{2}$,
and Keitaro Takahashi$^{3}$\\
$^{1}$Department of Physics, Kyoto University,
Kyoto 606-8502, Japan\\
$^{2}$Department of Physics, Faculty of Science,
Kanazawa University, Kakuma, Kanazawa, Ishikawa 920-1192, Japan\\
$^{3}$Yukawa Institute for Theoretical Physics, Kyoto University, 
Kyoto 606-8502, Japan}

\begin{document}


\pagerange{\pageref{firstpage}--\pageref{lastpage}} \pubyear{2008}

\maketitle

\label{firstpage}

\begin{abstract}

We extend the Hubble diagram up to $z = 5.6$ using 63 gamma-ray
bursts (GRBs) via peak energy-peak luminosity relation
(so called Yonetoku relation), and obtain constraints on
cosmological parameters including dynamical dark energy
parametrized by $P/\rho\equiv w(z) = w_0 + w_a \cdot z/(1+z)$.
It is found that the current GRB data are consistent with
the concordance model,
($\Omega_m = 0.28, \Omega_{\Lambda} = 0.72, w_0 = -1, w_a = 0$),
within two sigma level. Although constraints from GRBs
themselves are not so strong, they can improve the conventional
constraints from SNeIa because GRBs have much higher redshifts.
Further we estimate the constraints on the dark-energy
parameters expected by future observations with GLAST
(Gamma-ray Large Area Space Telescope) and \swift by
Monte-Carlo simulation. Constraints would improve
substantially with another 150 GRBs.

\end{abstract}


\section{Introduction}
\label{sec:intro}

In our previous paper \citep{Kodama2008}, we calibrated
the peak energy-peak luminosity relation of GRBs
with 33 nearby events ($z < 1.62$) whose luminosity distances
were estimated from those of large amount of SNeIa
\citep{Riess2007,Wood-Vasey2007,Davis2007}. This calibrated
Yonetoku relation, derived without assuming any cosmological
models, can be used as a new cosmic distance ladder toward
higher redshifts. Then we determined the luminosity distances
of 30~GRBs in $1.8 < z < 5.6$ using the calibrated relation
and calculated the likelihood varying
$(\Omega_m, \Omega_{\Lambda})$. We obtained
$(\Omega_m, \Omega_{\Lambda}) =
(0.37^{+0.14}_{-0.11}, 0.63^{+0.11}_{-0.14})$ 
for a flat universe, which is consistent with the concordance
cosmological model within one sigma level. Our logic to obtain
a new distance ladder is similar to that for SNeIa, that is,
we calibrate a new distance indicator (Yonetoku relation)
at low redshifts ($z < 1.62$) using the well established
indicators (SNeIa). Then we assume the new relation holds
at high redshifts ($z > 1.62$), although more detailed
analysis is needed for possible selection bias and evolution
effects in the relation \citep{Oguri2006}.

Currently the number of GRBs with $z > 1.62$ is relatively
small ($\sim 30$) and the statistical and systematic errors
of the Yonetoku relation are not so small compared with
SNeIa. Even then, GRBs are still effective to probe dark
energy, especially for dynamical dark energy whose energy
density becomes large at high redshifts, because the mean
redshift of GRBs is higher than that of SNeIa
\citep{Amati2008,Liang2008,Basilakos2008,
Schaefer2007,Ghirlanda2006,Firmani2006a}.
There would be many ways to characterize the time variation of
dark energy. Here we adopt a simple phenomenological model
as \citep{Chevallier2001,Linder2003}
$P/\rho \equiv w(z) = w_0 + w_a \cdot z/(1+z) = w_0 + w_a (1-a)$,  
where $w_0$ and $w_a$ are constants and $a$ is the scale factor
of the universe. For this model, GRBs would give strong
constraints on $w_a$ which represents the time dependence
of dark energy.

In this Letter, we present constraints on cosmological 
parameters such as $w_0$ and $w_a$ using both SNeIa with
$z < 1.8$ and GRBs with $z < 5.6$. In \S~2, we briefly review
the main results of the previous paper \citep{Kodama2008}
and the theoretical and observational basis of the analysis.
In \S~3-1, we assume cosmological constant ($w = -1$) and
obtain constraints on ($\Omega_m, \Omega_{\Lambda}$).
In \S~3-2, we assume a flat universe and non-dynamical
($w_a = 0$) dark energy and obtain constraints on
$(w_0, \Omega_m)$. In \S~3-3, we fix $\Omega_m = 0.28$ for
simplicity, and obtain the plausible values of $w_0$ and $w_a$. 
In \S~4, we discuss how these constraints will be improved
in future observations of high redshift GRBs by such as GLAST.
Throughout the paper, we fix the current Hubble parameter
as $H_0 = 66~{\rm km~s^{-1}Mpc^{-1}}$.

\section{Friedmann universe}

In our previous paper \citep{Kodama2008}, we calibrated the
Yonetoku relation \citep{Yonetoku2004} using 33~events with
the redshift $z < 1.62$ as
\begin{eqnarray}
\label{eq:ep-l}
(\frac{L_p}{10^{52}{\rm erg~s^{-1}}})
= (1.31 \pm 0.67)\times 10^{-4} 
(\frac{E_p}{\rm 1keV})^{1.68 \pm 0.09},
\end{eqnarray}
where $L_p$ and $E_p$ are the peak luminosity and 
the peak energy of the spectrum of a certain GRB event
in a comoving frame, respectively. Fig. 1 of the previous
paper shows that the linear correlation coefficient of
the above relation in logarithmic scale is 0.9478 and
the chance probability is $6.0 \times 10^{-17}$.
However the data distribution has a larger deviation
around the best fit line compared with the expected
Gaussian distribution. We estimated this systematic 
deviation in the normalization as $9.57 \times 10^{-5}$.

Now if one assumes that the above relation holds even for
$z > 1.8$, we can determine the luminosity distance $d_L(z)$
from the observed peak flux ($f_{p, {\rm obs}}$) and
the observed peak energy of the spectrum $E_{p,{\rm obs}}$
since $L_p = 4\pi d_L(z)^2 f_{p,{\rm obs}}$ and
$E_p = (1+z) E_{p,{\rm obs}}$. We express unknown equation
of state of dark energy as $P = w_X(z) \rho$ and
$\Omega_X$ as the present energy density of the dark energy
divided  by the critical density. Then in the Friedmann
universe with
$\Omega_k \equiv \Omega_m + \Omega_X - 1$,
the luminosity distance is given by 
\begin{eqnarray}
&& {d_L^{\rm th}}(z, \Omega_m, \Omega_X, w)\! \nonumber \\
&& ~~~ = \! \left\{
\begin{array}{ll}
\frac{c}{H_0 \sqrt{\Omega_k}} \sin(\sqrt{\Omega_k}F(z))
& \mbox{if}~~\Omega_k > 0\\
\frac{c}{H_0 \sqrt{-\Omega_k}} \sinh(\sqrt{-\Omega_k}F(z))
& \mbox{if}~~\Omega_k < 0\\
\frac{c}{H_0} F(z)
& \mbox{if}~~ \Omega_k = 0\\
\end{array}
\right.
\end{eqnarray}
with
\begin{eqnarray}
F(z)
&=& \int_0^z dz' \,
    \Big[\Omega_m (1+z')^3 - \Omega_k (1+z')^2
\nonumber \\
& & + \Omega_X
      e^{3 \int_0^{\ln(1+z')} d\ln(1+z'') [1+w(z'')]}
    \Big]^{-1/2}.
\label{F(z)}
\end{eqnarray}
Using the distance modulus ($\mu_0(z_i)$) from observation
of SNeIa and GRBs, we define a likelihood function as 
\begin{equation}
\Delta \chi^2
= \sum_{i}
  \left\{
  \frac{\mu_0({z_i}) - \mu^{\rm th}(z_i, \Omega_m, \Omega_X, w)}
       {\sigma_{\mu_{0, z_i}}}
  \right\}^2
  - \chi_{\rm best}^2,
\label{eq:contour1}
\end{equation}
where
$\mu^{\rm th}(z_i, \Omega_m, \Omega_X,w)
= 5 \log( d_L^{\rm th}/{\rm Mpc}) + 25$
and $\chi_{\rm best}^{2}$ represents the chi-square value for
the best fit parameter set of $\Omega_m$, $\Omega_X$ and $w$.

\begin{figure}
\includegraphics[width=84mm]{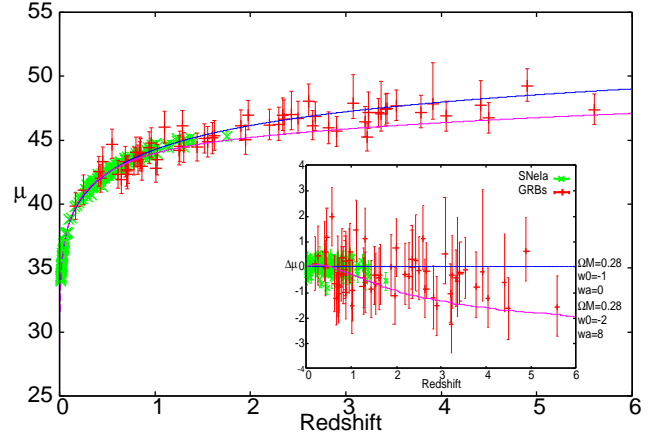}
 \vspace{0pt}
 \caption{Hubble diagram extended to $z = 5.6$ from GRBs.
 The green points and red points are the luminosity distance
 determined by SNeIa \citep{Riess2007} and GRBs \citep{Kodama2008},
 respectively. The blue line is the luminosity distance of
 $\Lambda$CDM model with ($\Omega_m$,$\Omega_{\Lambda}$)=(0.27,0.73).
 Inset figure is residual Hubble diagram and models after subtracting
 model ($\Omega_m,$,$\Omega_{\Lambda}$)=(0.27,0.73). The pink line
 is the luminosity distance of the dynamical dark energy equation
 of state model with ($w_0$, $w_a$)=(-2, 8) discussed in \S~3-3}
\label{fig1}
\end{figure}

\section{cosmological parameters}

Fig.~\ref{fig1} shows the Hubble diagram extended to $z = 5.6$ by
GRBs. The green points and red points are the luminosity distance
determined by SNeIa \citep{Riess2007,Wood-Vasey2007,Davis2007}
and GRBs \citep{Kodama2008}, respectively. The blue and pink lines
are the luminosity distances of $\Lambda$CDM model with
($\Omega_m$, $\Omega_{\Lambda}$)=(0.27, 0.73) and dynamical
dark energy model with ($w_0$, $w_a$)=(-2, 8), respectively.

\subsection{$\Lambda$CDM model}

\begin{figure}
\includegraphics[width=84mm]{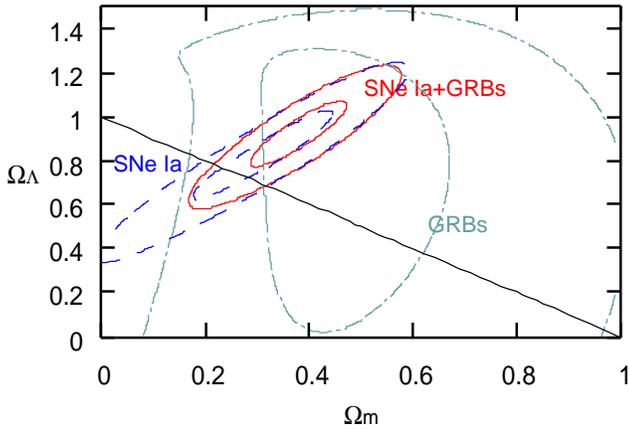}
 \vspace{0pt}
 \caption{The contours of likelihood $\Delta \chi^2$ in
 ($\Omega_m \- \Omega_{\Lambda}$) plane for GRBs
 (light blue dash-doted lines), SNeIa(blue dotted lines),
 SNeIa + GRBs (red solid lines), respectively. The contours
 correspond to 68.3\% and 99.7\% confidence regions,
 respectively, and black solid line represents the flat universe.
 The shape of GRB contour is more vertical because GRBs have
 higher redshifts.}
\label{fig2}
\end{figure}

We first consider the cosmological constant model,
that is  $w_0=-1$ and $w_a=0$. Then Eq.~(\ref{F(z)}) becomes
\begin{equation}
F(z) =
\int_0^z
\frac{dz}
     {\sqrt{\Omega_m (1+z)^3 + \Omega_{\Lambda} - \Omega_k (1+z)^2}},
\label{F(z)_lambda}
\end{equation}
and $\chi^2$ is a function of ($\Omega_m,\Omega_{\Lambda}$). 
Fig.~\ref{fig2} shows confidence regions for
($\Omega_m,\Omega_{\Lambda}$) from 63 GRBs
(light blue dash-dotted lines), 192 SNeIa (blue dotted lines),
and 63 GRBs + 192 SNeIa (red solid lines), respectively.
Without any prior the set of the cosmological parameters
with the largest likelihood is
($\Omega_m$,$\Omega_{\Lambda}$) =
($0.38_{-0.09}^{+0.09}$,$0.93_{-0.15}^{+0.14}$)
with $\chi_{\rm best}^2=225.2/253$.

\subsection{non-dynamical dark energy model}
\begin{figure}
\includegraphics[width=84mm]{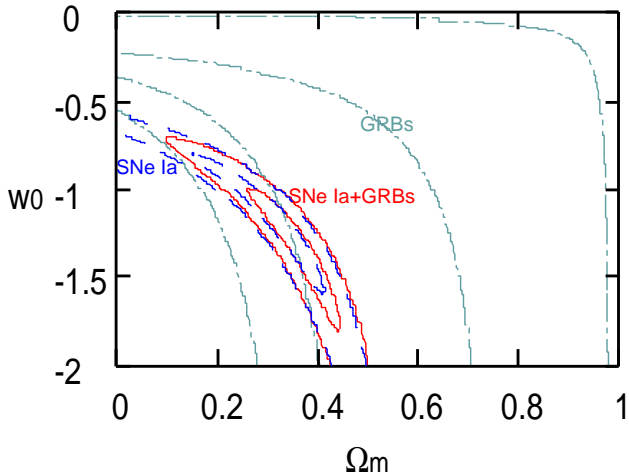}
 \vspace{0pt}
 \caption{The contours of likelihood of $\Delta \chi^2$ in
 ($\Omega_m$, $w_0$) plane for GRBs (light blue dash-dotted lines),
 SNeIa (blue dotted lines), SNeIa + GRBs (red  solid lines),
 respectively. The contours correspond to 68.3\% and 99.7\%
 confidence regions, respectively.}
\label{fig3}
\end{figure}

In this section we assume a flat universe with $w_a = 0$.
Then
\begin{equation}
F(z) =
\int_0^z
\frac{dz}{\sqrt{\Omega_m (1+z)^3 + (1-\Omega_m) (1+z)^{3(1+w_0)}}},
\label{F(z)_w}
\end{equation}
and the likelihood function depends on $\Omega_m$ and $w_0$.
Fig.~\ref{fig3} shows the likelihood contours on ($\Omega_m, w_0$)
plane for GRBs (light blue dash-dotted lines), SNeIa
(blue dotted lines), SNeIa + GRBs (solid red lines), respectively.
The contours correspond to 68.3\% and 99.7\% confidence regions,
respectively. The set of the cosmological parameters with the
largest likelihood is ($\Omega_m$, $w_0$)
= ($0.36_{-0.11}^{+0.08}$, $-1.33_{-0.15}^{+0.48}$)
with $\chi_{\rm best}^2 = 227.0/253$.

\subsection{Dynamical dark energy model}

\begin{figure}
\includegraphics[width=84mm]{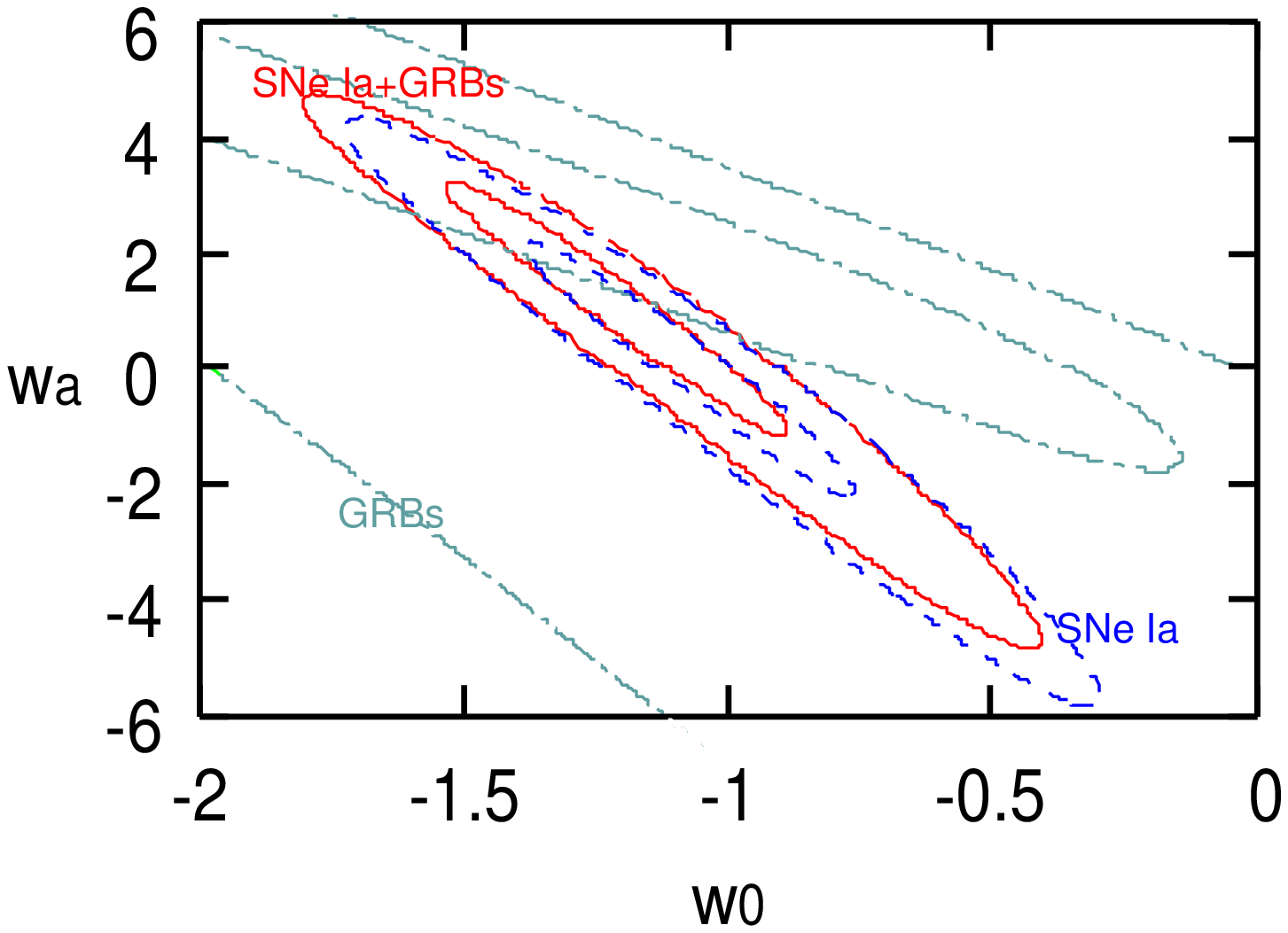}
 \vspace{0pt}
 \caption{The contours of likelihood of $\Delta \chi^2$ in 
 ($w_0$,$w_a$) plane for GRBs (light blue dash-doted lines),
 SNeIa (blue dotted lines), SNeIa + GRBs (red solid lines),
 respectively. The contours correspond to 68.3\% and 99.7 \%
 confidence regions, respectively.}
\label{fig4}
\end{figure}

As already shown, we adopt the parameterization of $w(z)$ as
\citep{Chevallier2001,Linder2003}
\begin{equation}
w(z)=w_0+w_a(1-a)=w_0+w_a\frac{z}{1+z}.
\end{equation}
This is not the only parameterization and, for example,
$w(z) = w_0 + w_1 z$ can be found in the literature.
However, $w(z)$ is diverging for large $z$ in this ($w_0, w_1$)
parameterization, which would not be appropriate for
our high redshift GRB samples. Now Eq.~(\ref{F(z)}) becomes
\begin{eqnarray}
F(z)
&=& \int_0^z dz \, \Big[\Omega_m(1+z)^3  \nonumber \\
& & (1 - \Omega_m) (1+z)^{3 (1 + w_0 + w_a)}
              e^{- 3 w_a \frac{z}{1+z}}
    \Big]^{-1/2}.
\label{F(z)_dynamical}
\end{eqnarray}
For simplicity, we fix $\Omega_m = 0.28$. Fig.~\ref{fig4} shows
the contours of likelihood $\Delta \chi^2$ in the ($w_0$, $w_a$)
plane from GRBs (light blue dash-dotted lines), SNeIa
(blue dotted lines), SNeIa + GRBs (red solid lines),
respectively. The contours correspond to 68.3\% and 99.7\%
confidence regions, respectively. The shape of probability
contour is more horizontal than that of SNeIa, because of
the higher redshift distribution of GRBs. The set of the
dark-energy parameters with the largest likelihood is 
($w_0$, $w_a$) = ($-1.26_{-0.32}^{+0.38}, 1.4_{-0.2}^{+1.8}$)
with $\chi^2_{\rm best} = 227.6/253$. The pink line in the
inset of Fig. \ref{fig1} is the model with
($w_0, w_a$) = ($-2, 8$) which is three sigma level from
the best fit. We can see that this model does not fit
the data by eyes.

Figs.~\ref{fig2}, \ref{fig3} and \ref{fig4} show that
the constraints on cosmological parameters from SNeIa are
stronger than those from GRBs at present. However the
shapes of the likelihood contours are different.
In Figs.~\ref{fig2} and \ref{fig3}, the contours from GRBs
are more vertical than that from SNeIa. This behaviour is
clear from Eqs. (\ref{F(z)_lambda}) and (\ref{F(z)_w}).
GRBs include higher redshifts up to $z = 5.6$ at present
so that the value of $F(z)$ is more sensitive
on the value of $\Omega_m$. Therefore we can expect
independent stronger constraints on cosmological parameters
if the systematic error in Yonetoku relation decreases
and/or the number of high redshift GRBs increases.
Fig.~\ref{fig4} shows also that the contour from GRBs
is more horizontal than SNeIa. Since the mean value of the
redshift for SNeIa samples and GRB samples are 0.48 and 1.97,
respectively, we see that GRBs should give more stronger
constraints on $w_a$ from the exponential term in
Eq. (\ref{F(z)_dynamical}).

\section{Future prospect of gamma-ray burst cosmology}

In this section we investigate the future prospect of probing
dark-energy parameters with GRBs. The Gamma-ray Large Area Space
Telescope (GLAST) was launched June 11, 2008 and it would
substantially increase the potential of GRBs as cosmological
probes. In fact, due to the wide energy-band of GLAST Burst
Monitor (GBM) and positional accuracy of Large Area Telescope
(LAT), GLAST is expected to detect 30 GRBs/year with spectral
peak energy and spectroscopic redshift by joint observation
with \swift.

Here we estimate the accuracy of determination of the dark-energy
parameters with GLAST by Monte-Carlo simulation.
We generate 150 GRB events in the following way. First,
GRBs are distributed in redshift-luminosity plane according
to the GRB formation rate and luminosity function given by
\citet{Porciani2001}. Spectral peak energy is assigned to each
GRB according to the best-fit Yonetoku relation with intrinsic
dispersion of $30 \%$. Calculating the flux at the earth assuming
the concordance model, a GRB is counted as an observed event
if the flux exceeds the sensitivity of GLAST. For observed events,
observational errors of $20 \%$ are added to the observed flux
and spectral peak energy. In this way, we generate 150 observed
events and they are divided into two groups, high-redshift group
(77 GRBs with $z > 1.7$) and low-redshift group
(73 GRBs with $z < 1.7$). As we did with real GRB events,
we reconstruct Yonetoku relation with low-redshift GRBs.
With increased number of low-redshift events, the normalization
and index of Yonetoku relation are determined with reduced errors
of $5 \%$ and $3 \%$, respectively. Applying the reconstructed
relation to high-redshift events, we can put them in the Hubble
diagram and constrain the cosmological parameters. For the details
of our Monte-Carlo method, see \citep{Takahashi2003,Oguri2006}.

\begin{figure}
\includegraphics[width=84mm]{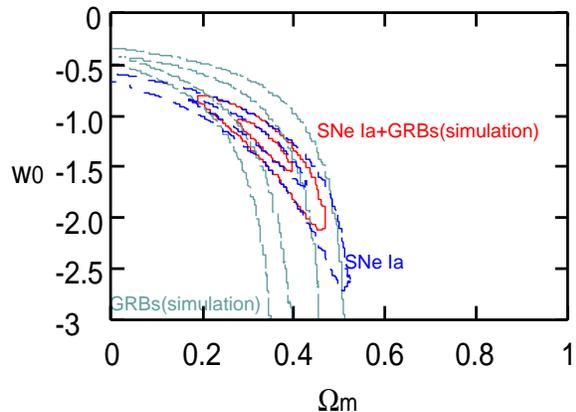}
 \vspace{0pt}
 \caption{The contours of likelihood of $\Delta \chi^2$ in
 ($\Omega_m$,$w_0$) plane from the real + simulated 
 GRBs (light blue dash-dotted lines), SNeIa (blue dotted lines),
 and SNeIa + GRBs (red solid lines), respectively. The contours
 correspond to 68.3\% and 99.7 \% confidence regions,
 respectively.}
\label{fig5}
\end{figure}

\begin{figure}
\includegraphics[width=84mm]{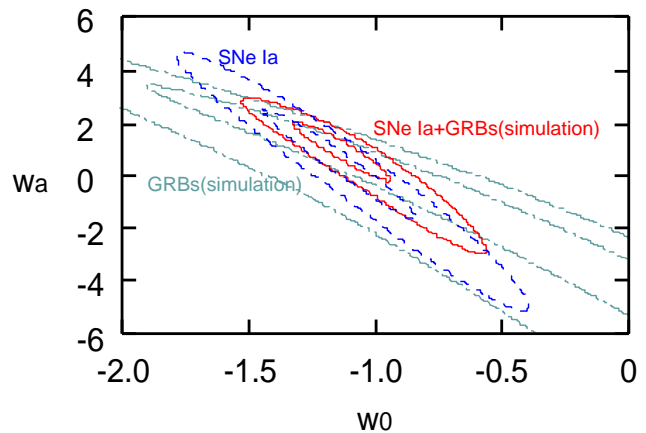}
 \vspace{0pt}
 \caption{The contours of likelihood of $\Delta \chi^2$ in
 ($w_0$,$w_a$) plane from the real + simulated
 GRBs (light blue dash-dotted lines), SNeIa (blue dotted lines),
 and SNeIa + GRBs (red solid lines), respectively. The contours
 correspond to 68.3\% and 99.7 \% confidence regions,
 respectively.}
\label{fig6}
\end{figure}

Figs.~\ref{fig5} and \ref{fig6} show the constraint in
($\Omega_m, w_0$) and ($w_0, w_a$) planes, respectively, from
the ``real + simulated'' high-redshift GRBs (light blue dash-dotted
lines), ``real'' SNeIa (blue dotted lines), and SNeIa + GRBs
(red solid lines). The contours correspond to 68.3\% and
99.7 \% confidence regions, respectively. These figures indicate
how GRBs can be a powerful probe to study the nature of dark
energy. 

In conclusion, the increase of high redshift GRB data by such as
GLAST and \swift is
indispensable to determine the time variation of the dark energy
in $z > 1.8$  where SNeIa data would be rare.

\section*{Acknowledgments}

This work is supported in part by the Grant-in-Aid from the 
Ministry of Education, Culture, Sports, Science and Technology
(MEXT) of Japan,  No.19540283,No.19047004, No.19035006(TN),
and  No.18684007 (DY) and  by the Grant-in-Aid for the global
COE program
{\it The Next Generation of Physics, Spun from Universality and Emergence}
from MEXT of Japan. KT is supported by a Grant-in-Aid
for the Japan Society for the Promotion of Science (JSPS) Fellows and
is a research fellow of  JSPS.

\end{document}